\documentclass[english,british,letter]{aa}
\usepackage[T1]{fontenc}
\usepackage[latin9]{inputenc}
\usepackage{verbatim}
\usepackage{amsmath}
\usepackage{graphicx}
\usepackage[authoryear]{natbib}

\makeatletter

\providecommand{\tabularnewline}{\\}
\newcommand{\lyxdot}{.}

\usepackage{txfonts}

\def\Myr{\mathrm{Myr}}
\def\Gyr{\mathrm{Gyr}}
\def\enp{\enlargethispage{2.5cm}}

\usepackage{babel}
\makeatother

\begin{document}

\title{Merger as Intermittent Accretion}

\author{Morgan Le Delliou\inst{1,2}}

\institute{Instituto de Física Teórica \\
Módulo C-XI Facultad de Ciencias Universidad Autónoma de Madrid \\
Cantoblanco, 28049 Madrid SPAIN\\
\email{\selectlanguage{english}%
Morgan.LeDelliou@uam.es\selectlanguage{british}
} \and Centro de F\'{\i}sica Teórica e Computacional, Universidade
de Lisboa \\
 Av. Gama Pinto 2, 1649-003 Lisboa, Portugal\\
\email{\selectlanguage{english}%
delliou@cii.fc.ul.pt\selectlanguage{british}
}}

\date{Received...; Accepted...\hfill{}Preprint: DF-IFT/UAM--08-13\\
 \phantom{m}\hfill{}astro-ph:0705.1144v2\vspace{-0.5cm}
}

\abstract{} {The Self-Similar Secondary Infall Model (SSIM) is modified to
simulate a merger event.} {The model encompasses spherical versions
of tidal stripping and dynamical friction that agrees with the Syer
\& White merger paradigm's behaviour.} {The SSIM shows robustness
in absorbing even comparable mass perturbations and returning to its
original state.} { It suggests the approach to be invertible and
allows to consider accretion as smooth mass inflow merging and mergers
as intermittent mass inflow accretion. }

\keywords{Cosmology:theory -- Dark Matter -- Galaxies:formation -- galaxies:halos
-- gravitation}

\maketitle
%
{}\vspace{-0.5cm}

\section{Introduction\label{sec:Introduction}}

\enp Structure formation in the Cold Dark Matter (CDM, or more simply
DM) paradigm is dominated by the hierarchical picture of repeated
mergers. This picture was emphasised by \citet{Syer&White}, explaining
the dynamical formation of halo density profile with a feedback mechanism
provided by repeated mergers. Whereas it is now believed that isotropisation
of the velocity dispersion \citep[angular momentum; see ][]{LeDH03,BarnesEtal05,MacMillanEtal06}
via the radial-orbit instability (also viewed as adiabatic variability
of self-similarity, \citealt{Hcg07}) is responsible for the density
profile formation, their picture remains a widely accepted description
of the merger digestion mechanism. Despite its simple spherical symmetry
and apparent lack of compliance with the merger paradigm, some studies
have shown that the Secondary Infall Model (SIM) is a viable model
to predict the structure and density profile evolutions of DM haloes
as compared to N-body simulations \citep{Ascasibar07,SalvadorEtal07}. 

This letter proposes to understand this paradox by examining the merger
paradigm within the SIM and studying how merger events impact on the
relaxation and structure of a CDM halo. 

The SIM stems from the seminal work of \citet{GunnGott}, and the
SSIM (Self-similar SIM) started when \citet{FG84} and \citet{Bertschinger85}
independently found self-similar solutions to the SIM. It was later
shown that those solutions can be reached from non-self-similar initial
conditions \citep[e.g. in][]{HoffShah85,WhiteZaritsky,Ryden93,HW95,HW97,AvilaReese99,HW99,delPopolo00,HLeD02,LeDH03}
and a systematic approach to the SSIM was used in \citet{HW95,HW97,HW99,HLeD02,LeDH03},
derived from the Carter-Henriksen formalism \citep[hereafter CH]{CH91}.
Some extensions to the SIM were proposed that included the effects
of angular momentum to explain flat halo cusps \citep{Hioletis02,LeDH03,Arascibar04,Williams04,Lu06},
but no fundamental attempt was made before \citet{MLeDPhD02} to confront
the SIM with the merger paradigm.

The following section (Sec. \ref{sec:Merger-in-an}) will describe
how and why the SSIM can be extended to model a merger event. Then
Sec. \ref{sec:Merger-paradigm-and} will discuss how the symmetry
of the SSIM still allows for a form of tidal stripping and dynamical
friction, before presenting the consequences of such a merger in the
SSIM in Sec. \ref{sec:Digestions}, and to make some concluding remarks
in Sec. \ref{sec:Discussion-and-conclusions}.

\section{Merger in an Infall\label{sec:Merger-in-an}}

\enp Modelling a merger event in a spherical geometry may appear
contradictory but it is possible to a certain extent. To understand
this it is important to realise the following: a very small amount
of substructures are seen in N-body simulations;  \citet{Diemand07}
find that only 5.3\% of the total mass fraction of haloes lie in subhaloes.
In the \citet{Syer&White} picture, incoming satellite haloes merge
with their parent, fall in the centre and contribute to the density
profile and to the parent's relaxation and virialisation. However,
in simulations, subobjects swing back and forth several times in their
parents before being digested. That process can be modelled in a simpler
way: on average, spherical symmetry is not bad \citep{Ascasibar07}
as it reproduces the correct time scales and density profiles. Shell
codes are much \textbf{simpler} than N-body codes and therefore provide
with robust tests of certain aspects of their results. Other simplifying
approaches have been used to understand halo formation, such as phase-space
coarse graining \citep{LeDH03,Hcg04,Hcg06} or in the one dimensional
slab model used in \citet{Binney04}, where it was shown to explain
the formation of cosmic web sheets through the interplay of phase
mixing and violent relaxation, also present in spherical models. \citet{HW99}
have shown that relaxation is moderately violent (in their figure
9) and induced by a phase space instability \citep{HW97}. Section
\ref{sec:Merger-paradigm-and} will detail how another perspective
of phase mixing and moderately violent relaxation through phase space
instability can be interpreted as some sort of tidal stripping and
dynamical friction.

In this paper the SSIM is implemented in a shell code (see details
in \citealp{MLeDPhD02}, and references therein) with fully dynamical
Lagrangian treatment of infall using the CH \citep{CH91} self-similar
variables that reveals when the system reaches naturally a self-similar
regime. A halo is modelled from a radial power law perturbation $\delta\rho/\rho\propto r^{-\epsilon}$
on an Einstein-de Sitter homogeneous background, that is evolved to
reach its quasi-stationary self-similar regime in its core%
\footnote{The core, or self gravitating system, is defined as the set of shells
in the multiple flow region. Its edge's radius is that of the outermost
shell that has passed only once through the centre, as seen in phase
space.%
} \citep{HW99}. The SIM is known to establish a self-similar infall
phase \citep{HW97}, which then leads to a semi-universal power law
density profile \citep{FG84,Bertschinger85}: for initial power index
$\epsilon\le2$, the isothermal sphere ($\rho\propto r^{-\mu}$ with
$\mu=2$) is the semi-universal attractor, whereas with $\epsilon>2$,
there is a continuum of attractors with $\mu=3\epsilon/(1+\epsilon)$.
Positive overdensity and the requirement of a finite initial core
mass in the centre limit the range to $0\le\epsilon<3$. The cores
explored here were chosen, as presented in Table \ref{tab:ODinputParams},
according to their SSIM behaviour defined by their initial power index:
typical shallow ($\epsilon=3/2$) and steep ($\epsilon=5/2$) profiles,
with the addition of an extreme steep case ($\epsilon=2.9$) to test
the behaviour of a highly concentrated parent halo. The steep and
shallow denominations refer to the comparison relative to the isothermal
sphere. 

\begin{table}
\begin{centering}
\begin{tabular}{cccc}
\hline 
$\epsilon$, panel & $M_{ratio}$ & $D_{ratio}$ & $M_{OD}/M_{BG}$\tabularnewline
\hline
$3/2,$ upper panel & 0.751 & 0.282 & 1.173\tabularnewline
$3/2,$ middle panel & 4.25$\times10^{-2}$ & 7.10$\times10^{-2}$ & 9.38$\times10^{-2}$\tabularnewline
$3/2,$ lower panel & 6.92$\times10^{-2}$ & 0.168 & 1.453\tabularnewline
\hline
$5/2,$ upper panel & 0.889 & 5.51$\times10^{-2}$ & 0.319\tabularnewline
$5/2,$ middle panel & 0.439 & 5.54$\times10^{-2}$ & 0.290\tabularnewline
$5/2,$ lower panel & 0.178 & 0.454 & 1.133\tabularnewline
\hline
$2.9,$ upper panel & 0.753 & 9.19$\times10^{-2}$ & 0.416\tabularnewline
$2.9,$ middle panel & 0.407 & 0.641 & 1.118\tabularnewline
$2.9,$ lower panel & 0.301 & 9.71$\times10^{-2}$ & 0.344\tabularnewline
\hline
\end{tabular}
\par\end{centering}

\caption{\label{tab:ODinputParams}Density, mass and mass perturbation ratios
defining the satellite initial OD for the mergers in the SSIM. The
mass perturbation measures the perturbation of the OD compared to
the background halo region it spans, just before entering the core.
First column gives parent initial power law seed and panel order in
reference to figures \ref{fig:VirialODPhSp1.5}, \ref{fig:VirialODPhSp2.5}
and \ref{fig:VirialODPhSp2.9}.}

\end{table}
In this geometry, an overdensity (hereafter OD, or satellite), representing
a spherically averaged satellite halo, is a region of overdense shells
close to the edge of the core, the parent halo (hereafter core, or
parent). 

\enp The OD is evolved dynamically from an initial gau\ss ian density
profile added on top of the background halo profile over a finite
region. That evolution runs long enough to observe the signature of
the OD's own stationary regime in phase space. This is manifested
in the mixing of its Liouville sheet during the OD's dynamical mass
accretion of halo shells from its environment. The OD's definition
as a set of particles (shells) is frozen when the core swallows it. 

At that point are recorded the ratios of OD-over-core masses, $M_{ratio}$,
of their densities, $D_{ratio}$, and the measure of the perturbation
provided by the OD on its background surroundings, in mass, $M_{OD}/M_{BG}$.
For each case, three different satellites were chosen, trying to obtain
various types of mass and density ratios between satellites and parents. 

Since they were allowed to accrete mass dynamically from their environment,
ODs were laid close to the edge of the core to maintain some control
over the final frozen mass and density ratios. Some configurations
of those ratios were too difficult to obtain: in the shallow case,
with high $M_{ratio}$, lower values for $D_{ratio}$ were prevented
by the high density background the OD accretes from, while for the
steep cases, also with high $M_{ratio}$, higher $D_{ratio}$ couldn't
be obtained because of their cores' poor density backgrounds which
tended to spread the ODs (see Sec. \ref{sec:Digestions}'s tidal effect).

The ratios indicated are measured at the time of core entry. The explored
values are presented in Table \ref{tab:ODinputParams}. 

It is crucial to point out that the numerical implementation of the
SSIM entails a shell code where finite size shells model the continuous
system. That will play a role in the discussion of the results.

\section{Merger paradigm and SSIM\label{sec:Merger-paradigm-and}}

\begin{figure}
\begin{centering}
\includegraphics[width=1\columnwidth,height=0.4\textwidth,keepaspectratio]{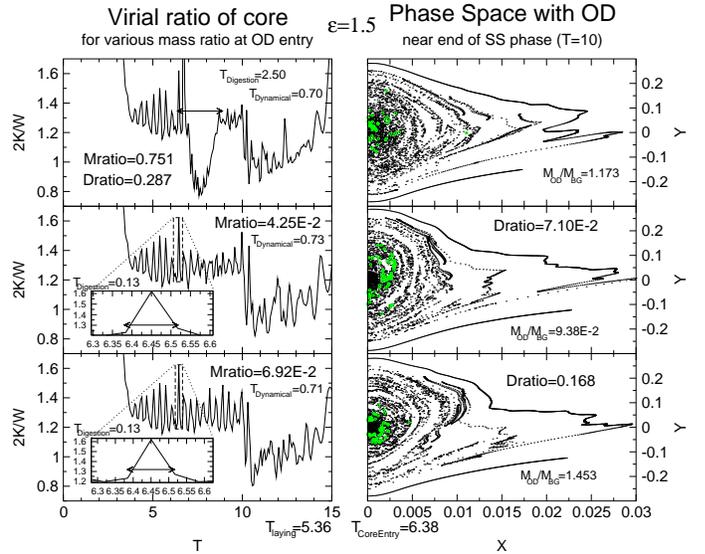}
\par\end{centering}

\caption{\label{fig:VirialODPhSp1.5}Shallow case: Virial ratio and phase space
diagrams at the end of the self-similar phase for three sets of ODs
in the $\epsilon=\frac{3}{2}=1.5$ case. Overdensity shells are emphasised
in green in phase space. Digestion time is defined from OD core entry
(pre-spike, see Sec. \ref{sec:Digestions}) to virial recovery (measured
on left panels). Zoomed encapsulation of those spikes in middle and
lower left panels show their measure in troughless cases. T, X and
Y are respectively the self-similar time, radius and radial velocity,
which units are set by $G=M(\infty)=1$ \citep{HW97}.}

\end{figure}
\enp \citet{Syer&White} have attempted to define the singularity
of mergers in an effort, at the time, to explain the universality
of the density profile found in N-body simulation by \citet[hereafter NFW]{NFW96}:
their key feature is the feedback mechanism between dynamical friction
from the parent halo and tidal stripping of the satellite. Even though
this is not anymore considered to hold the key to the formation of
the density profile, their merger digestion mechanisms is still widely
accepted to describe the behaviour of satellites. I argue that both
mechanisms can be modelled within the SSIM despite its one-dimensional
nature.

Tidal acceleration on an infinitesimal shell of mass $dm=4\pi\rho r^{2}dr$
-- located at radius $r$, containing the system mass $M$ and with
thickness $dr$ -- can be defined as the differential gravity between
its boundaries. Defining the cumulative average density profile\begin{align}
\left\langle \rho\right\rangle _{r}= & \frac{M(r)}{4\pi r^{3}/3},\end{align}
the inward oriented elementary tidal acceleration reads, to leading
order,\begin{align}
dT= & 4\pi Gdr\left(\rho-\frac{2}{3}\left\langle \rho\right\rangle _{r}\right).\end{align}
It is thus clear that regions of peak density below the cumulative
average ($\rho<\frac{2}{3}\left\langle \rho\right\rangle _{r}$) will
experience a net disruptive tidal acceleration spreading apart shells
in those regions, in the radial direction. In this spherically averaged
study of a merger, this models tidal stripping.

Dynamical friction classically is defined as the creation of a wake
by a moving mass in a gravitating medium which back reaction entails
a net drag force upon the moving mass. In the SSIM, a massive shell
is crossing the core's shell in its travelling inwards or outwards.
We will see that a radial drag force, with the correct orientation,
is experienced as a result of this motion in the spherically averaged
model. 

This crossing of shells by the OD's results in shells just outside
of it feeling more or less mass pulling inwards, depending on the
direction of the motion of the massive OD shells. That leads to a
differential tightening or spreading of core's shell behind the moving
mass, in the fashion of a wake. However in \emph{spherical symmetry},
an outer wake \textbf{does not} contribute to the pull on the OD.
Nevertheless, its mass corresponds to shells that \emph{defected}
from inside because of OD motion, and their effect can be seen in
the dynamics (see Appendix \ref{sec:Spherical-model-of}).

In a similar fashion, the dynamical effect on the OD from its motion
can be \emph{described} in terms of a drag force: the crossing of
core shells by massive OD shell lead to a decrease, or increase, of
the resulting inner mass of the moving OD, depending on the direction
of motion. Thus, with inner mass goes the inner pull, which can be
interpreted a dragging force that adds to the total force, that should
be experienced in the opposite direction of the motion.

\enp Therefore, the SSIM with an outer overdensity can be interpreted
to model the main features of the merger paradigm.

\section{Digestions\label{sec:Digestions}}

\begin{figure}
\begin{centering}
\includegraphics[width=1\columnwidth,height=0.4\textwidth,keepaspectratio]{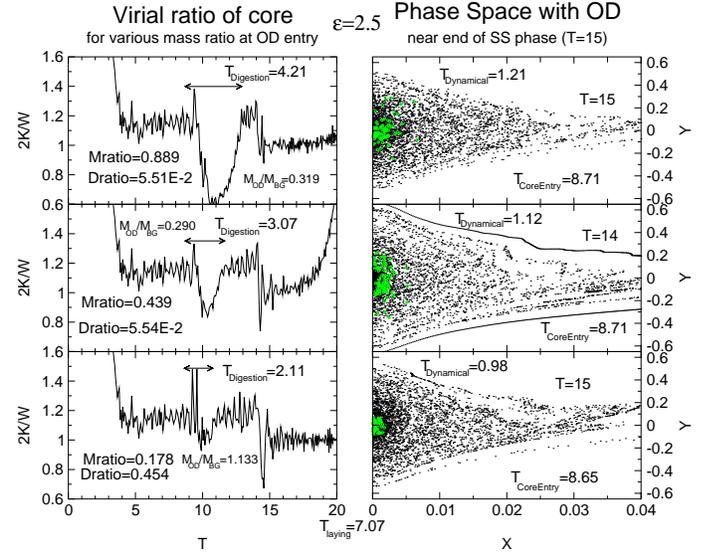}
\par\end{centering}

\caption{\label{fig:VirialODPhSp2.5}Steep case: Virial ratio and phase space
diagrams at the end of the self-similar phase, for three sets of ODs
in the $\epsilon=\frac{5}{2}=2.5$ case, including an emphasis on
digested overdensity shells in phases space and a measure of digestion
time. Same units as in Fig. \ref{fig:VirialODPhSp1.5}.}

\end{figure}

\begin{figure}
\begin{centering}
\includegraphics[width=1\columnwidth,height=0.4\textwidth,keepaspectratio]{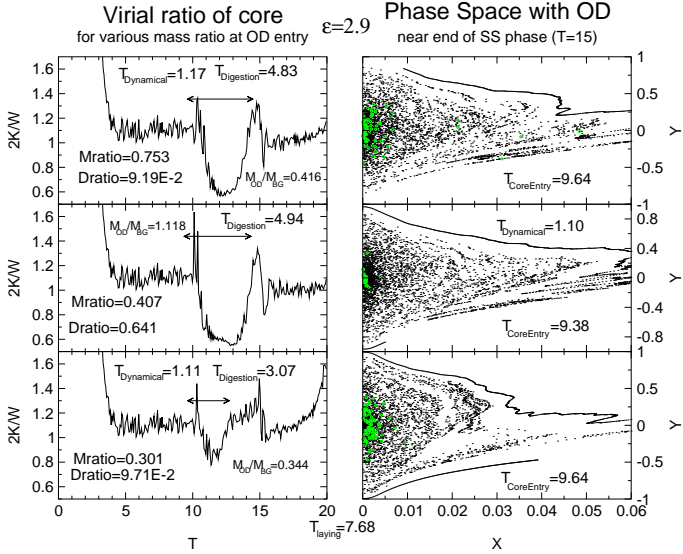}
\par\end{centering}

\caption{\label{fig:VirialODPhSp2.9}Extreme steep case: Virial ratio and phase
space diagrams at the end of the self-similar phase, for three sets
of ODs in the $\epsilon=2.9$ case, including an emphasis on digested
overdensity shells in phase space and a measure of digestion time.
Same units as in Fig. \ref{fig:VirialODPhSp1.5}.}

\end{figure}

\begin{table}
\begin{centering}
\begin{tabular}{ccccc}
\hline 
$\epsilon$, panel & $T_{digestion}$ & $T_{dynamical}$ & $\frac{T_{digestion}}{T_{dynamical}}$ & $M_{ratio}.D_{ratio}$\tabularnewline
\hline 
$\frac{3}{2},$ upper p. & 2.50 & 0.70 & 3.57 & 0.212\tabularnewline
$\frac{3}{2},$ middle p. & 0.13 & 0.73 & 0.178 & 3.017$\times10^{-3}$\tabularnewline
$\frac{3}{2},$ lower p. & 0.13 & 0.71 & 0.183 & 1.163$\times10^{-2}$\tabularnewline
$\frac{5}{2},$ upper p. & 4.21 & 1.21 & 3.48 & 4.989$\times10^{-2}$\tabularnewline
$\frac{5}{2},$ middle p. & 3.07 & 1.12 & 2.74 & 2.432$\times10^{-2}$\tabularnewline
$\frac{5}{2},$ lower p. & 2.11 & 0.98 & 2.15 & 8.081$\times10^{-2}$\tabularnewline
$2.9,$ upper p. & 4.83 & 1.17 & 4.13 & 6.920$\times10^{-2}$\tabularnewline
$2.9,$ middle p. & 4.94 & 1.10 & 4.49 & 2.609$\times10^{-1}$\tabularnewline
$2.9,$ lower p. & 3.07 & 1.11 & 2.77 & 2.923$\times10^{-2}$\tabularnewline
\hline
\end{tabular}
\par\end{centering}

\caption{\label{tab:ODoutputParams}Digestion and dynamical times and strength
parameter of the OD for the mergers in the SSIM. Again, first column
gives parent initial power law seed and panel order in figures \ref{fig:VirialODPhSp1.5},
\ref{fig:VirialODPhSp2.5} and \ref{fig:VirialODPhSp2.9}.}

\end{table}
Indeed, it is possible to keep track, in the Lagrangian shell model,
of the defined satellite's (OD's) components once they have been absorbed
by the parent (core). The core can be considered isolated at the end
of the accretion phase \citep{HW97}. The phase space configurations
of simulated merged haloes are displayed on Figs. \ref{fig:VirialODPhSp1.5},
\ref{fig:VirialODPhSp2.5}, and \ref{fig:VirialODPhSp2.9}'s right
panels, distinguishing between the core and OD's accreted shells.
This reveals how the different ODs, in their various (shallow or steep)
environments, either retain some degree of coherence after being ingested
by the core or have been digested and scattered over the core's phase
space. 

The left panels of Figs. \ref{fig:VirialODPhSp1.5}, \ref{fig:VirialODPhSp2.5},
and \ref{fig:VirialODPhSp2.9} examine the Virial ratios of the corresponding
cores, and show a remarkable robustness in the SSIM: the quasi-stable
self-similar phase%
\footnote{with Virial markedly different from usual value of 1!%
} is shown to be either marginally or strongly disturbed by the OD
absorption, but to return to the \textbf{original} undisturbed level
of the parent after a digestion time $T_{digestion}$, provided a
mass flow still fuels the self-similar equilibrium. Digestion is manifested
by a more or less pronounced initial decrease (entry of extra mass
in core increases W), followed by a spike (first crossing of centre
gives $m_{OD}$ high velocities, thus peaks K) and then, for stronger
disturbance, a trough (energy exchanges from phase space instability,
shells spend on average more time at low velocities, thus lower Virial,
\citealp{HW99}). Its deepness depends primarily on $M_{ratio}$.
Digestion time measurements are shown on Figs. \ref{fig:VirialODPhSp1.5},
\ref{fig:VirialODPhSp2.5}, and \ref{fig:VirialODPhSp2.9}'s left
panels (double horizontal arrows), and are summarised in Table \ref{tab:ODoutputParams}.
There, they are compared with the OD's free fall dynamical time through
the core, $T_{dynamical}$, also indicated on the figures. $T_{dynamical}$
is defined as the free fall time to the centre of a test shell across
a constant density distribution, equivalent to the core, in self-similar
variables. From Table \ref{tab:ODoutputParams}, without Fig. \ref{fig:VirialODPhSp1.5}'s
two lowest panels, where the definition of $T_{digestion}$ is problematic,
the average $\left\langle T_{digestion}/T_{dynamical}\right\rangle =3.33$,
with a standard deviation of $0.77$, can be computed. It shows the
core digests the OD in 2 to 4 passages in the central relaxation region
of phase space. This is comparable to the number of distinguishable
Lagrange-Liouville streams present in the core's outer phase space
regions, as seen from Figs. \ref{fig:VirialODPhSp1.5}, \ref{fig:VirialODPhSp2.5},
and \ref{fig:VirialODPhSp2.9}'s right panels.

\enp From the OD's point of view, the mergers display their effects
in phase spaces, represented on  Figs. \ref{fig:VirialODPhSp1.5},
\ref{fig:VirialODPhSp2.5}, and \ref{fig:VirialODPhSp2.9}'s right
panels, on which two features are crucial: the spread (or compactness)
of the OD over the core at the end of the infall phase and the presence
of some, or all, of its shells in the centre of the core's phase space.
This reflects the digestion mechanisms adopted by \citet{Syer&White}.
Their proposal aimed at a dynamical explanation of the NFW profile.
Although this explanation is not anymore considered (see Sec. \ref{sec:Introduction}),
it is interesting to note that the presently discussed single merger
model in the SSIM shows signs of inflections (central flattening and
edge steepening) from its semi-universal, almost isothermal, density
profile. However this is not the focus of this paper.

The OD's compactness  resists to tidal stripping while its final presence
in the centre is driven by dynamical friction. The fate of a model
satellite in the SSIM displays behaviour well in agreement with the
merger digestion mechanisms proposed by \citeauthor{Syer&White}:
in the SSIM a combination of density and mass ratios leads to emphasise
each effect. High $D_{ratio}$s seem to be the dominant factor for
OD's compactness, while high $M_{ratio}$s promote the sinking of
the OD to the centre of the core's phase space. 

All possible qualitative types of behaviour are present: if both ratios,
$M_{ratio}$ and $D_{ratio}$, are strong enough, the OD survives
almost intact to the centre of phase space (Figs. \ref{fig:VirialODPhSp2.5}'s
lower and \ref{fig:VirialODPhSp2.9}'s middle right panels). If only
$M_{ratio}$ is high while $D_{ratio}$ is low, the OD is scattered
at the centre (Figs. \ref{fig:VirialODPhSp1.5}, \ref{fig:VirialODPhSp2.5}
and \ref{fig:VirialODPhSp2.9}'s upper right panels). Conversely,
a high $D_{ratio}$ and low $M_{ratio}$ lead to a compact OD around,
but not reaching, the centre of phase space (Fig. \ref{fig:VirialODPhSp1.5}'s
lower right panel). %
{} Finally if both ratios are too low, the OD is scattered without reaching
the centre of phase space (Figs. \ref{fig:VirialODPhSp1.5} and \ref{fig:VirialODPhSp2.5}'s
middle and \ref{fig:VirialODPhSp2.9}'s lower right panels). 

A step further in this phenomenology would be to note that a combination
of both ratios should be taken ($M_{ratio}.D_{ratio}$, see Table
\ref{tab:ODoutputParams}), for which a threshold can be defined for
reaching the centre and another for compactness of the OD. However
this classification seems to require an additional dependency with
the steepness of the initial profile. Indeed the available data offer
different ranges for each initial profile case. The shallow case calls
for higher values for the $M_{ratio}.D_{ratio}$ thresholds than the
steep cases. This reflects the shallow case's wider spread of material,
compared with the steep cases, that the OD has to cross on its journey
towards the centre of phase space.%

\enp As an illustration of our model, we can assume the Milky Way
(hereafter MW) to have a shallow profile and use the corresponding
reliable digestion time model, that is with $\epsilon=1.5$, $M_{ratio}=0.751$
and $T_{digestion}=2.50$. The corresponding satellite S would have
a mass $M_{S}\simeq44M_{LMC}$ compared to the Large Magellanic Cloud
(hereafter LMC), which is huge. The model then yields a very short
digestion time, also compared with the age of the oldest stars in
the MW $T_{MW}=13.2\Gyr$, as\begin{align}
T_{digestion}\simeq & 584\Myr\simeq\frac{T_{MW}}{22.6}.\end{align}
Its dynamical time $T_{dynamical}\simeq234\Myr$ indicates that at
the end of digestion, this satellite's shells would be lined between
the second incoming and second outgoing inner streams of the core
and the model suggests it to then sink to the centre by the end of
the MW formation as seen on Fig. \ref{fig:VirialODPhSp1.5}'s upper
right panel.

\section{Discussion and conclusions\label{sec:Discussion-and-conclusions}}

The SSIM has proven its capacity to model a merger event. Its simplicity
allows one to probe the dynamics of the merger and the most remarkable
result of this work shows that the self-similar quasi-stable regime
of quasi-Virial equilibrium is extremely \textbf{robust} to perturbations
that can be of comparable size to the core (equal mass mergers): the
Virial ratio, after a more or less long period of digestion returns
to its stabilised \textbf{original} undisturbed level, after only
2 to 4 passages in the centre, and continues its usual evolution.
The spreading and sinking of the satellite's particles across the
parents and towards its centre agree with the tidal stripping and
dynamical friction picture from \citet{Syer&White}, provided some
adaptation to the language of the SSIM's symmetry. Finally, and this
is the claim of this paper, the numerical implementation of the model
requiring discretisation, the rapid oscillations of the Virial ratio
in the accretion phase offer a novel interpretation in the light of
the SSIM merger model: instead of a continuous stream of mass, the
model presents a repeated bombardment of finite mass shells that can
be understood as small overdensities; Fig. \ref{fig:VirialODPhSp1.5}'s
zoomed two lowest right panels show a spike to manifest the weakest
mergers digestion; thus the wiggles in the Virial ratio can be interpreted
as manifestation of repeated mergers that are at this level indistinguishable
from accretion. Therefore there is \emph{no fundamental difference}
between mergers and accretion, the latter being a series of repeated
merger with vanishing mass, while the latter is just intermittent
accretion. This reconciles approaches such as \citet{SalvadorEtal07}
where accretion was presented as a memory loss mechanism, eliminating
the need to refer to mergers.\enp 

\begin{acknowledgements}
The work of MLeD is supported by CSIC (Spain) under the contract JAEDoc072,
with partial support from CICYT project FPA2006-05807, at the IFT,
Universidad Autónoma de Madrid, Spain, and was also supported by FCT
(Portugal) under the grant SFRH/BD/16630/2004,  at the CFTC, Lisbon
University, Portugal. Evidently, thanks should go the R.N.Henriksen
for discussions and comments and for directing MLeD's thesis work
from which these results are extracted.
\end{acknowledgements}
\bibliographystyle{aa}
\bibliography{ODpaper}

\begin{thebibliography}{30}
\expandafter\ifx\csname natexlab\endcsname\relax\def\natexlab#1{#1}\fi

\bibitem[{Ascasibar {et~al.}(2007)Ascasibar, Hoffman, \&
  Gottlöber}]{Ascasibar07}
Ascasibar, Y., Hoffman, Y., \& Gottlöber, S. 2007, MNRAS, 376, 393

\bibitem[{Ascasibar {et~al.}(2004)Ascasibar, Yepes, Gottlöber, \&
  Müller}]{Arascibar04}
Ascasibar, Y., Yepes, G., Gottlöber, S., \& Müller, V. 2004, MNRAS, 352, 1109

\bibitem[{Avila-Reese {et~al.}(1999)Avila-Reese, Firmani, \&
  Klypin}]{AvilaReese99}
Avila-Reese, V., Firmani, C., \& Klypin, A.and~Kravtsov, A. 1999, MNRAS, 310,
  527

\bibitem[{Barnes {et~al.}(2005)Barnes, Williams, Babul, \&
  Dalcanton}]{BarnesEtal05}
Barnes, E., Williams, L., Babul, A., \& Dalcanton, J. 2005, ApJ, 643, 797

\bibitem[{Bertschinger(1984)}]{Bertschinger85}
Bertschinger, E. 1984, ApJS, 58, 39

\bibitem[{Binney(2004)}]{Binney04}
Binney, J. 2004, MNRAS, 350, 939

\bibitem[{Carter \& Henriksen(1991)}]{CH91}
Carter, B. \& Henriksen, R. 1991, JMPS

\bibitem[{del Popolo {et~al.}(2000)del Popolo, Gambera, Recami, \&
  Spedicato}]{delPopolo00}
del Popolo, A., Gambera, M., Recami, E., \& Spedicato, E. 2000, A\&A, 353, 427

\bibitem[{Diemand {et~al.}(2007)Diemand, Kuhlen, \& Madau}]{Diemand07}
Diemand, J., Kuhlen, M., \& Madau, P. 2007, ApJ, 657, 262

\bibitem[{Fillmore \& Goldreich(1984)}]{FG84}
Fillmore, J. \& Goldreich, P. 1984, ApJ, 281, 1

\bibitem[{Gunn \& Gott(1972)}]{GunnGott}
Gunn, J. \& Gott, J. 1972, ApJ, 176, 1

\bibitem[{Henriksen(2004)}]{Hcg04}
Henriksen, R. 2004, MNRAS, 355, 1217

\bibitem[{Henriksen(2006)}]{Hcg06}
Henriksen, R. 2006, MNRAS, 366, 697

\bibitem[{Henriksen(2007)}]{Hcg07}
Henriksen, R. 2007, ApJ, 671, 1147

\bibitem[{Henriksen \& Le~Delliou(2002)}]{HLeD02}
Henriksen, R. \& Le~Delliou, M. 2002, MNRAS, 331, 423

\bibitem[{Henriksen \& Widrow(1995)}]{HW95}
Henriksen, R. \& Widrow, L. 1995, MNRAS, 276, 679

\bibitem[{Henriksen \& Widrow(1997)}]{HW97}
Henriksen, R. \& Widrow, L. 1997, Phys.Rev.Lett., 78, 3426

\bibitem[{Henriksen \& Widrow(1999)}]{HW99}
Henriksen, R. \& Widrow, L. 1999, MNRAS, 302, 321

\bibitem[{Hiotelis(2002)}]{Hioletis02}
Hiotelis, N. 2002, A\&A, 382, 84

\bibitem[{Hoffman \& Shaham(1985)}]{HoffShah85}
Hoffman, Y. \& Shaham, J. 1985, ApJ, 297, 16

\bibitem[{Le~Delliou(2002)}]{MLeDPhD02}
Le~Delliou, M. 2002, PhD thesis, Queen's University, Canada

\bibitem[{Le~Delliou \& Henriksen(2003)}]{LeDH03}
Le~Delliou, M. \& Henriksen, R. 2003, A\&A, 408, 27

\bibitem[{Lu {et~al.}(2006)Lu, Mo, Katz, \& Weinberg}]{Lu06}
Lu, Y., Mo, H., Katz, N., \& Weinberg, M. 2006, MNRAS, 368, 1931

\bibitem[{MacMillan {et~al.}(2006)MacMillan, Widrow, \&
  Henriksen}]{MacMillanEtal06}
MacMillan, J., Widrow, L., \& Henriksen, R. 2006, ApJ, 653, 43

\bibitem[{Navarro {et~al.}(1996)Navarro, Frenck, \& White}]{NFW96}
Navarro, J., Frenck, C., \& White, S. 1996, ApJ, 462, 563, (NFW)

\bibitem[{Ryden(1993)}]{Ryden93}
Ryden, B. 1993, ApJ, 418, 4

\bibitem[{Salvador-Solé {et~al.}(2007)Salvador-Solé, Manrique,
  Gonz\'alez-Casado, \& Hansen}]{SalvadorEtal07}
Salvador-Solé, E., Manrique, A., Gonz\'alez-Casado, G., \& Hansen, S. 2007,
  ApJ, 666, 181

\bibitem[{Syer \& White(1998)}]{Syer&White}
Syer, D. \& White, S. 1998, MNRAS, 293, 337

\bibitem[{White \& Zaritsky(1992)}]{WhiteZaritsky}
White, S. \& Zaritsky, D. 1992, ApJ

\bibitem[{Williams {et~al.}(2004)Williams, Babul, \& Dalcanton}]{Williams04}
Williams, L., Babul, A., \& Dalcanton, J. 2004, ApJ, 604, 18

\end{thebibliography}

\appendix

\section{Spherical model of dynamical friction\label{sec:Spherical-model-of}}

A thin OD shell in the inward/outward direction, adding mass $\pm dm_{OD}$,
crossing shells at $r$ creates a differential acceleration w.r.t.
the state without OD which induces an infinitesimal displacement,
thus a wake, \begin{align}
dr= & \mp\frac{G(dt)^{2}dm_{OD}}{2r^{2}}.\end{align}
This wake of mass $dm_{W}=\rho r^{2}dr$ induces on the OD an acceleration
(backreaction) \begin{align}
a_{drag}= & -\frac{Gdm_{W}}{r^{2}}=-G\rho dr=\pm\frac{(Gdt)^{2}dm_{OD}}{2r^{2}}\rho,\end{align}
opposite to the direction of motion. In addition, the amplitude of
the drag force is shown proportional to $dm_{OD}\rho$, related to
$M_{ratio}.D_{ratio}$.
\end{document}